\documentclass[usenatbib,usegraphicx]{mn2e} 

\usepackage{url}
\usepackage[breaklinks=true]{hyperref}
\usepackage{twoopt}

\usepackage{natbib}

\usepackage[utf8x]{inputenc}
\usepackage[normalem]{ulem}
\usepackage{siunitx}
\usepackage{amsmath, amssymb}
\usepackage[farskip=0pt]{subfig}

\usepackage{multirow,bigstrut,ctable}
\usepackage[normalem]{ulem}
\usepackage{rotating}
\usepackage[varg]{txfonts} 

\def \arcmin      {$^\prime$}
\def \arcsec      {$^{\prime\prime}$}

\def \mjybeam     {mJy\,beam$^{-1}$}
\def \mujybeam    {$\mu$Jy\,beam$^{-1}$}
\def \klambda     {k$\lambda$}
\newcommand{\hi}{\text{H\textsc{i}}}

\newcommand{\beam}[2]{{#1}\arcsec$\times${#2}\arcsec}

\def \vla        {\emph{VLA}}

\def \rosat      {\emph{ROSAT}}
\def \gmrt       {\emph{GMRT}}
\def \wsrt       {\emph{WSRT}}

\def \target       {NGC~5580/NGC~5588}
\def \stokesi      {Stokes~{\textit I}}

\title[Radio emission from NGC~5580 \& NGC~5588]
      {The diffuse radio emission around NGC~5580 \& NGC~5588}

\author[F.~de~Gasperin et~al.]{F. de Gasperin$^{1}$, H.T. Intema$^{2}$, W. Williams$^{3,4}$, M. Br\"uggen$^{1}$, M. Murgia$^{5}$, 
\newauthor R. Beck$^{6}$, A. Bonafede$^{1}$
\\
$^{1}$ Universit\"at Hamburg, Hamburger Sternwarte, Gojenbergsweg 112, D-21029, Hamburg, Germany\\
$^{2}$ National Radio Astronomy Observatory, 1003 Lopezville Road, Socorro, NM 87801-0387, USA\\
$^{3}$ Leiden Observatory, Leiden University, P.O.Box 9513, NL-2300 RA, Leiden, The Netherlands\\
$^{4}$ Netherlands Institute for Radio Astronomy (ASTRON), PO Box 2, 7990AA, Dwingeloo, The Netherlands\\
$^{5}$ INAF -- Osservatorio Astronomico di Cagliari, Loc. Poggio dei Pini, Strada 54, I-09012, Capoterra (CA), Italy\\
$^{6}$ Max-Planck-Institut for Radioastronomy, Auf dem H\"ugel 69, D-53121, Bonn, Germany}
\begin{document}

\date{}
\pagerange{\pageref{firstpage}--\pageref{lastpage}} \pubyear{2013}
\maketitle

\label{firstpage}

\begin{abstract}
The galaxy pair NGC~5580 and NGC~5588 are part of a loose group of galaxies. They are surrounded by steep-spectrum, extended radio emission which was previously suggested to be a down-scaled example of Mpc-size radio haloes present in galaxies clusters.

We present a multi-frequency study of the radio-emission aimed to clarify its nature. The source has been observed with the Giant Meterwave Radio Telescope (\gmrt{}) at 235, 325 and 610 MHz and the images obtained were combined with archival data to cover the frequency range 150--1400 MHz.

The new observations revealed the presence of a second, fainter lobe on the South-East of NGC~5580. The spectral index study of the source shows a flattening of the spectrum (which implies a younger particle population) close to the two galaxies. We argue that the extended radio emission is the remnant of a past activity cycle of the active galactic nucleus (AGN) present in NGC~5580 and therefore a notable example of a dying radio galaxy located outside a dense environments.
\end{abstract}


\section{Introduction}

The galaxy pair NGC~5580/NGC~5588 is surrounded by an unidentified, bright and steep-spectrum radio emission. These two galaxies form a loose group with NGC~5557 \citep[LGG~378;][]{Garcia1993}. Other publications define the group to include up to 13 galaxies \citep[e.g., G~141;][]{Geller1983}, while \cite{Fouque1992} group NGC~5590 with a few more galaxies (NGC~5557, NGC~5529 and UGC~9113; R7-A6) but exclude NGC~5588. The peculiar radio emission covering (and apparently connecting) the two galaxies has been so far ignored. Only \cite{Delain2006} noticed it in the WENSS survey and classified it as a ``galaxy group radio-halo''. However, the source is not an X-ray emitter ($\log L_X < 41.2$ erg/s) and the galaxies are part of a rather small ($\lesssim 13$ galaxies) and loose group with high velocity dispersion \citep[461 km/s;][]{VanDriel2001}. Here we present pointed radio observations aimed at understanding the nature of this emission which may be a down-sized version of the Mpc-scale radio halos present in massive clusters of galaxies \citep{Venturi2008, Bonafede2009, Feretti2012} or it may be related to past episodes of AGN activity \citep{Schoenmakers2000, Parma2007, Murgia2011}.

The active stage of an AGN can last several tens of Myrs. During this period the radio emission associated with the galaxy is powered by a continuous flow of energy from the AGN which reached distances several times the hosting galaxy size via plasma beams and jets. In galaxy clusters the injection of energy sustains the growth of the radio lobes against the pressure in the hot X-ray emitting external medium into which they expand. After this period the AGN activity ceases or is dramatically reduced to the point that the plasma outflow can no longer sustain the radio source, which undergoes a period of fading (dying phase). During this phase only the radio lobes remain visible. After a time that depends on the source environment, also the emission becomes too faint to be detected and the source vanishes.

Dying radio galaxies are an important but poorly explored stage of AGN evolution. It is believed that the low-frequency radio emission from the fading radio lobes lasts longer \citep[$\sim 10^8$ yr;][]{Murgia2011} if their expansion is prevented by the pressure of a dense intra-cluster medium (ICM). In this way cosmic ray electrons (CRe) lose their energy only through synchrotron radiation and inverse Compton scattering, while expansion plays only a marginal role. It is also possible for the source to restart before the emission from a previous burst faded completely \citep{Schoenmakers2000}. The best example of fossil radio lobes seen in a currently active galaxy is 3C 338 \citep[see e.g.,][]{Gentile2007} where steep-spectrum lobes of the source are clearly disconnected from the currently active jets. 

%

In this paper we study the radio emission surrounding the galaxy pair NGC~5580/NGC~5588. In Sec.~\ref{sec:obs} we discuss the set-up and the data reduction of the \gmrt{} observations, and the archival \vla{} data used. In Sec.~\ref{sec:images} and Sec.~\ref{sec:spidx} we analyse the images and the spectral behaviour of the source. In Sec.~\ref{sec:discussion} we discuss various explanations for the radio source.

In this paper we adopt a $\Lambda$CDM cosmology with $H_0 = 71$, $\Omega_M = 0.27$, $\Omega_{\Lambda} = 0.73$. NGC~5580 is at a redshift $z = 0.0107$ ($D_L=45.6$ Mpc), which gives a scale of 13 kpc/arcmin.

\section{Observations}\label{sec:obs}

Using the \gmrt{}\footnote{\gmrt{}: \url{http://gmrt.ncra.tifr.res.in}}, we observed the extended emission around \target{} during two distinct runs of 8 hours each. During the night of May 2--3, 2013, visibilities were recorded in the dual-frequency mode at 610 and 235~MHz simultaneously, each using a single polarization (RR and LL, respectively). During the afternoon and evening of July 11, 2013,  visibilities were recorded at 325~MHz in dual-polarization (RR and LL). In all cases we used an 8~second integration time, and a 33~MHz bandwidth divided into 512 channels. The (primary) flux- and bandpass calibrator 3C~295 was observed at the start and end of each run, while the (secondary) gain calibrator 3C~286 was observed every hour for ten minutes. The total time-on-target per observing run was 368 and 398~minutes, respectively.

\subsection{Data reduction}\label{sec:datared}

For each observing frequency, the visibility data was processed using two independent data reduction pipelines: one based on AIPS\footnote{AIPS: \url{http://www.aips.nrao.edu}} incorporating SPAM ionospheric calibration \citep{Intema2009}, and the second based on CASA\footnote{CASA: \url{http://casa.nrao.edu}} and the automated flagging package AOflagger \citep{Offringa2012}. The best images in terms of signal-to-noise ratio and fidelity of the extended structure were then chosen. Maybe not surprisingly, the AIPS pipeline yielded better results at 235 and 325 MHz due to ionospheric calibration, while the CASA pipeline performed better at 610~MHz due to CASA's improved multi-scale CLEAN image reconstruction.

\begin{figure}
\centering
\includegraphics[width=.85\columnwidth]{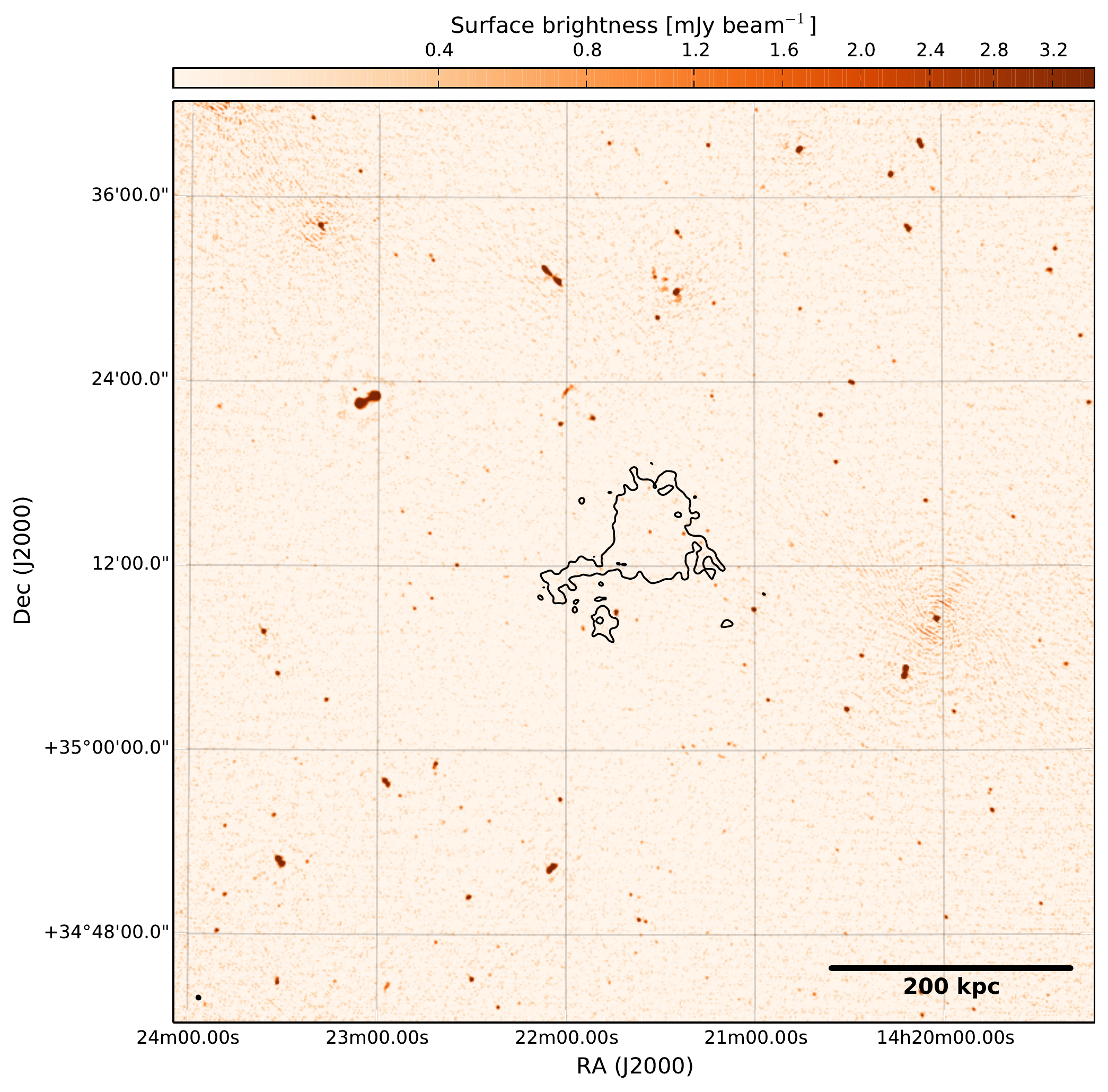}
\caption{The radio environment of \target{} at 323 MHz ($\text{rms} = 110$~\mujybeam). The contour shows the $3\sigma$ level (values in Table~\ref{tab:campaign}) of the low-resolution map.}\label{fig:330_HR}
\end{figure}

The AIPS-based, semi-automated pipeline processing is very similar for each of the \gmrt{} observing frequencies, being different only in the detailed settings of automated flagging. The pipeline uses the ParselTongue interface \citep{Kettenis2006} to access AIPS tasks, files and tables from Python. At 235 and 325 MHz, flux and bandpass calibrations were derived from 3C~286 after three iterations of flagging and calibrating. Additionally, instrumental phase calibrations were determined by filtering out ionospheric contributions \citep[see][]{Intema2009}, an important step for direction-dependent ionospheric calibration later on. Calibrations were transferred and applied to the target field data, simple clipping of spurious visibility amplitudes was applied, and data was averaged in time and frequency and converted to \stokesi{} to speed up processing. The effective bandwidths after flagging and averaging are 16.7 and 31.25~MHz, centered on 235 and 323~MHz, respectively. Self-calibration of the target field was started with an initial phase calibration using a multi-point source model derived from the NVSS, WENSS and VLSS radio source catalogs \citep{Condon1998, Rengelink1997, Cohen2007}, followed by (facet-based) wide-field imaging and CLEAN deconvolution of the primary beam area and several bright outlier sources (out to 5 primary beam radii). The visibility weighting scheme used during imaging is a mix between uniform and robust weighting (AIPS ROBUST -1), which generally provides a well-behaved point spread function (without broad wings) by downweighting the very dense central UV-coverage of the \gmrt{}. Self-calibration was repeated three more times, including amplitude calibration in the final round, and outlier flagging on residual (source-model-subtracted) visibilities in between imaging and calibration. Next, two rounds of SPAM calibration and imaging were performed to minimize direction-dependent, residual phase calibration errors due to the ionosphere. In each round, (i) bright sources in and around the primary beam area were peeled \citep[e.g.,][]{Noordam2004}, (ii) the peeling phase solutions were fitted with a time-variable two-layer phase screen model, (iii) the model was used to generate ionospheric phase correction tables for the grid of viewing directions defined by the facet centers, and (iv) the target field was re-imaged and deconvolved while applying the appropriate correction table per facet. We refer to \citet{Intema2009} for more details. The background rms noise in the images is 480~\mujybeam{} for a \beam{12}{11} beam at 235~MHz, and 110~\mujybeam{} for a \beam{11}{8.8} beam at 325~MHz (see Fig.~\ref{fig:330_HR}). For imaging the diffuse emission, we first created a high-resolution image by leaving out data with baselines shorter than 2~\klambda{}. After subtracting the high-resolution source model form the visibilities, we created a low-resolution image by leaving out data with baseline lengths longer than 8~\klambda{}, and simultaneously applying a circular Gaussian taper with the 30~percent point at 5~\klambda{}. For the resulting images, the background rms noise is 1.5~\mjybeam{} for a \beam{41}{23} beam at 235~MHz, and 380~\mujybeam{} for a \beam{30}{26} beam at 325~MHz. The resulting primary-beam corrected images are presented in Fig.~\ref{fig:240_LR} and Fig.~\ref{fig:330_LR}.

The CASA-based, semi-automated pipeline is written in Python. Data at 610 MHz were initially flagged through a series of increasingly sensitive runs of the CASA tasks FLAGDATA in the ``rflag'' mode. During these cycles a refined bandpass was obtained. Both 3C286 and 3C295 were used as flux and bandpass calibrator, for the latter a 2-point model had to be used at these frequencies, since the source was partially resolved. Flags were then reset and, once corrected for the bandpass, AOflagger \citep{Offringa2012} was used to detect and remove radio frequency interference. At the end of the procedure seven antennas were manually removed and around 25\% of the rest of the data was flagged because of interference. After an initial phase and amplitude calibration on 3C286, bad baselines were identified and removed by inspecting the BLCAL calibration tables, and the phase and amplitude solutions extended to the target field. The data of the target field were then SPLITted and averaged down to 32 channels, with a central frequency of 612 MHz and an effective bandwidth of 31.7 MHz. Several cycles of phase self-calibration, with subsequent clipping of the outliers, were performed and a final cycle of amplitude self-calibration was eventually made. Peeling of the brightest sources \citep[e.g.,][]{Noordam2004} is done iteratively, correcting in time/amplitude in their direction after subtracting the best model of the rest of the field. The source is then subtracted and the other sources restored. The resulting image has a background noise of 45~\mujybeam{} and a resolution of \beam{5}{4}. A high-resolution model of the field was extracted considering only baselines longer than 4~\klambda{}. This model was subtracted from the visibilities and a final image was made. For this last step, baselines longer than 8~\klambda{} were removed and data were tapered applying a circular Gaussian taper with FWHM at 5~\klambda{} to reach a resolution of \beam{31}{22} and enhance the extended emission of the target. The resulting primary-beam corrected image is presented in Fig.~\ref{fig:610_LR}.

\subsection{Archival data}\label{sec:archival}

\begin{figure*}
\centering
\subfloat[\gmrt{} 153 MHz.]{\includegraphics[width=.9\columnwidth]{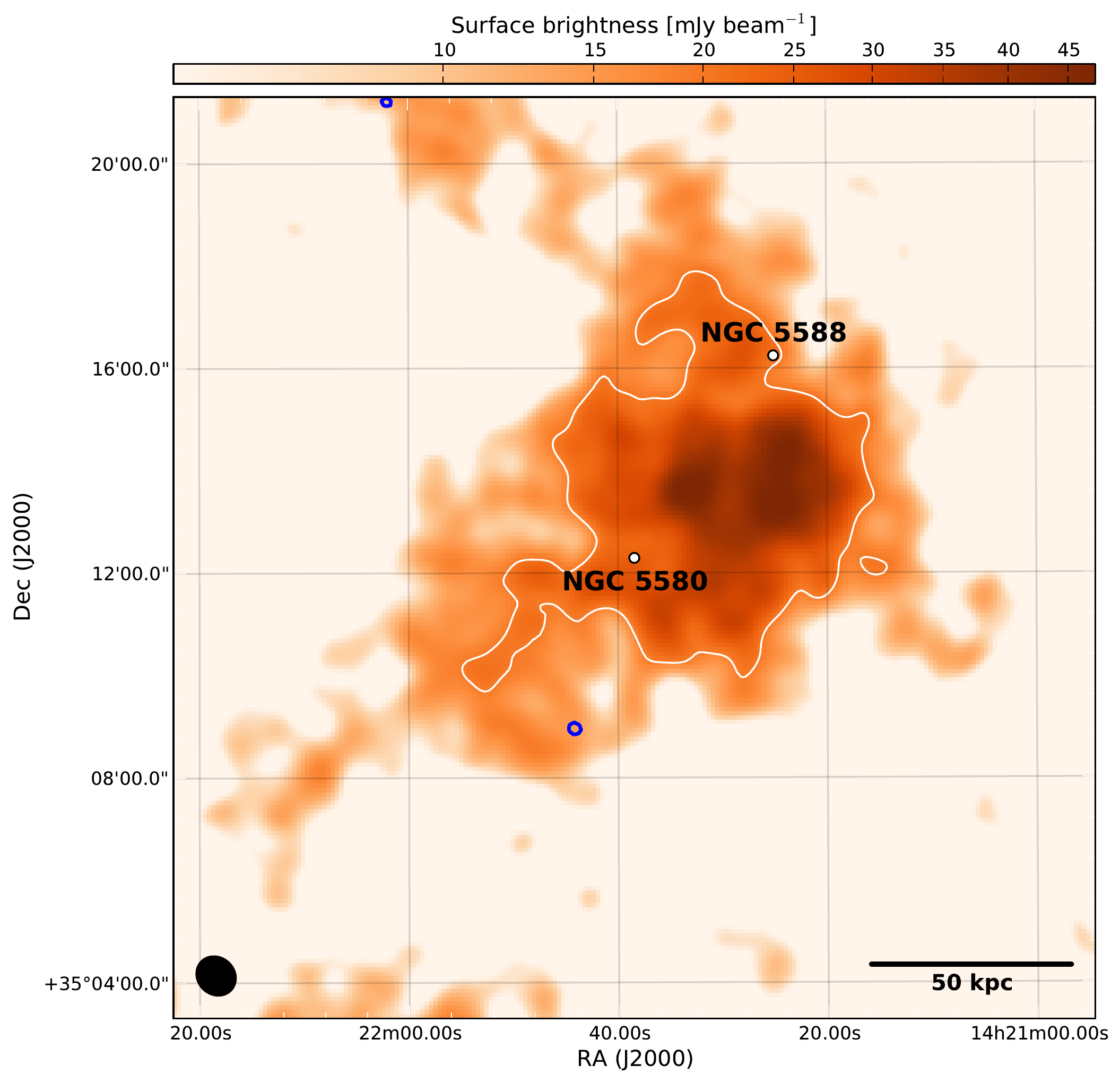}\label{fig:150_LR}}
\subfloat[\gmrt{} 235 MHz.]{\includegraphics[width=.9\columnwidth]{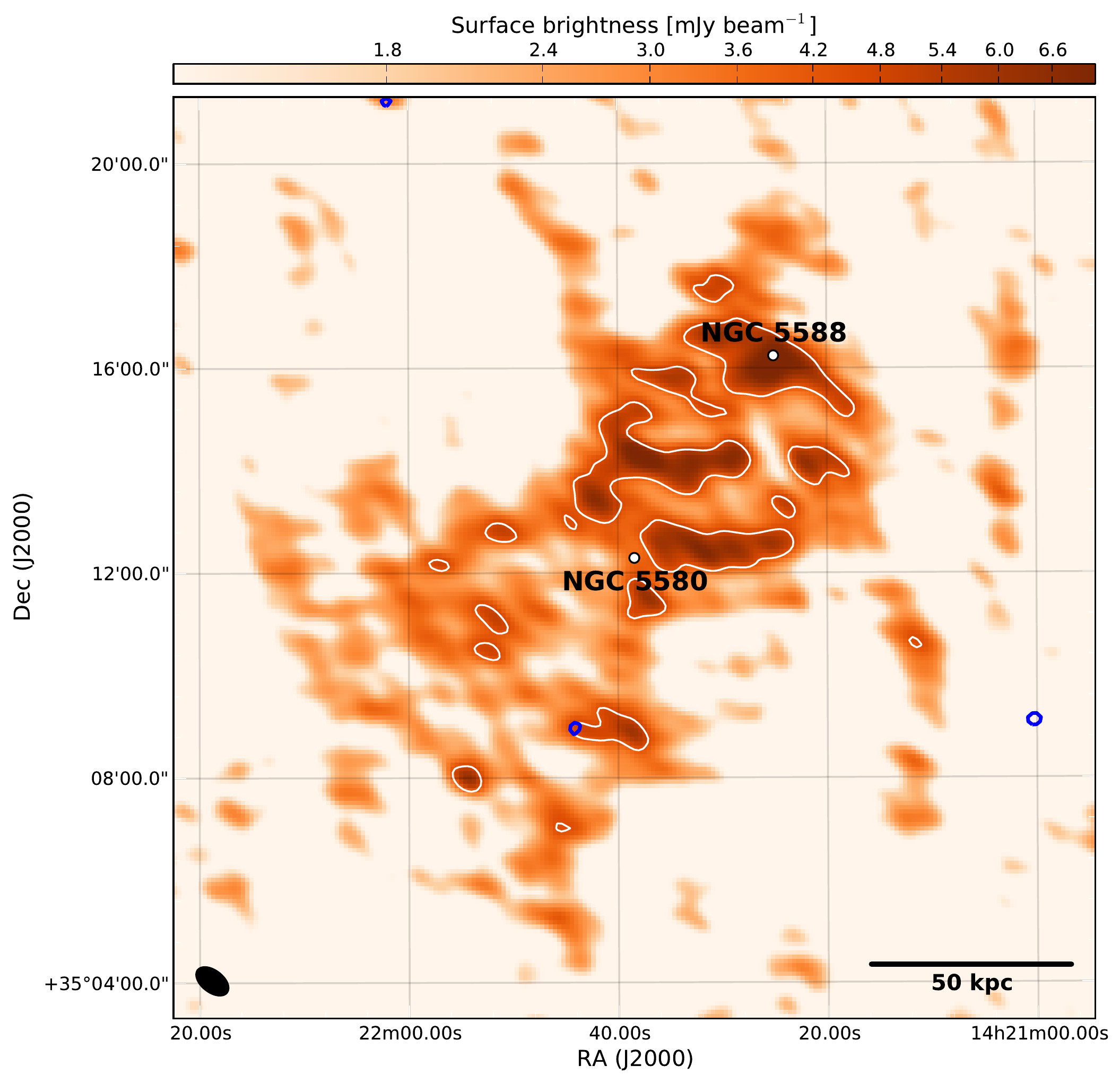}\label{fig:240_LR}}\\
\subfloat[\gmrt{} 323 MHz.]{\includegraphics[width=.9\columnwidth]{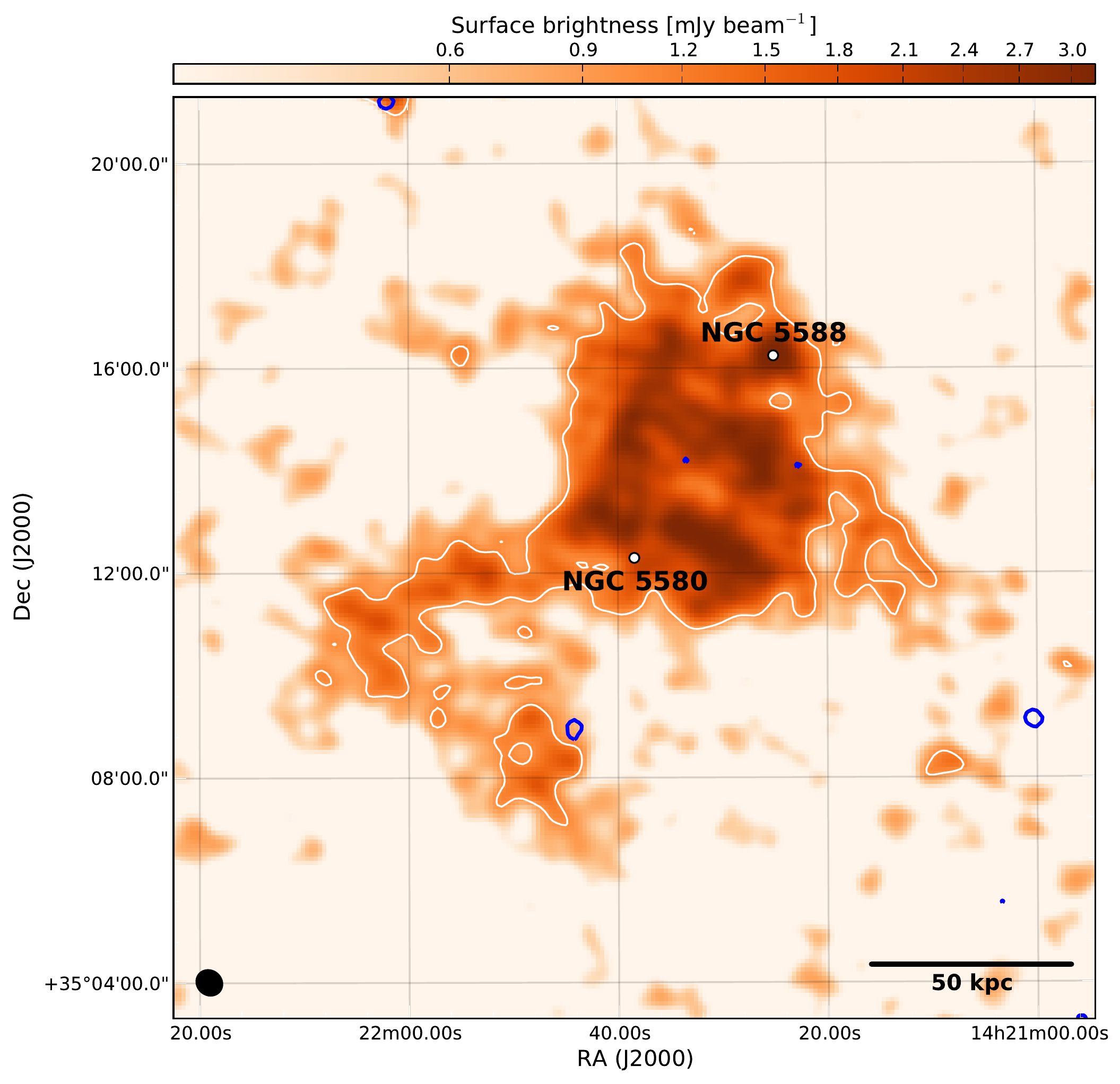}\label{fig:330_LR}}
\subfloat[\gmrt{} 610 MHz.]{\includegraphics[width=.9\columnwidth]{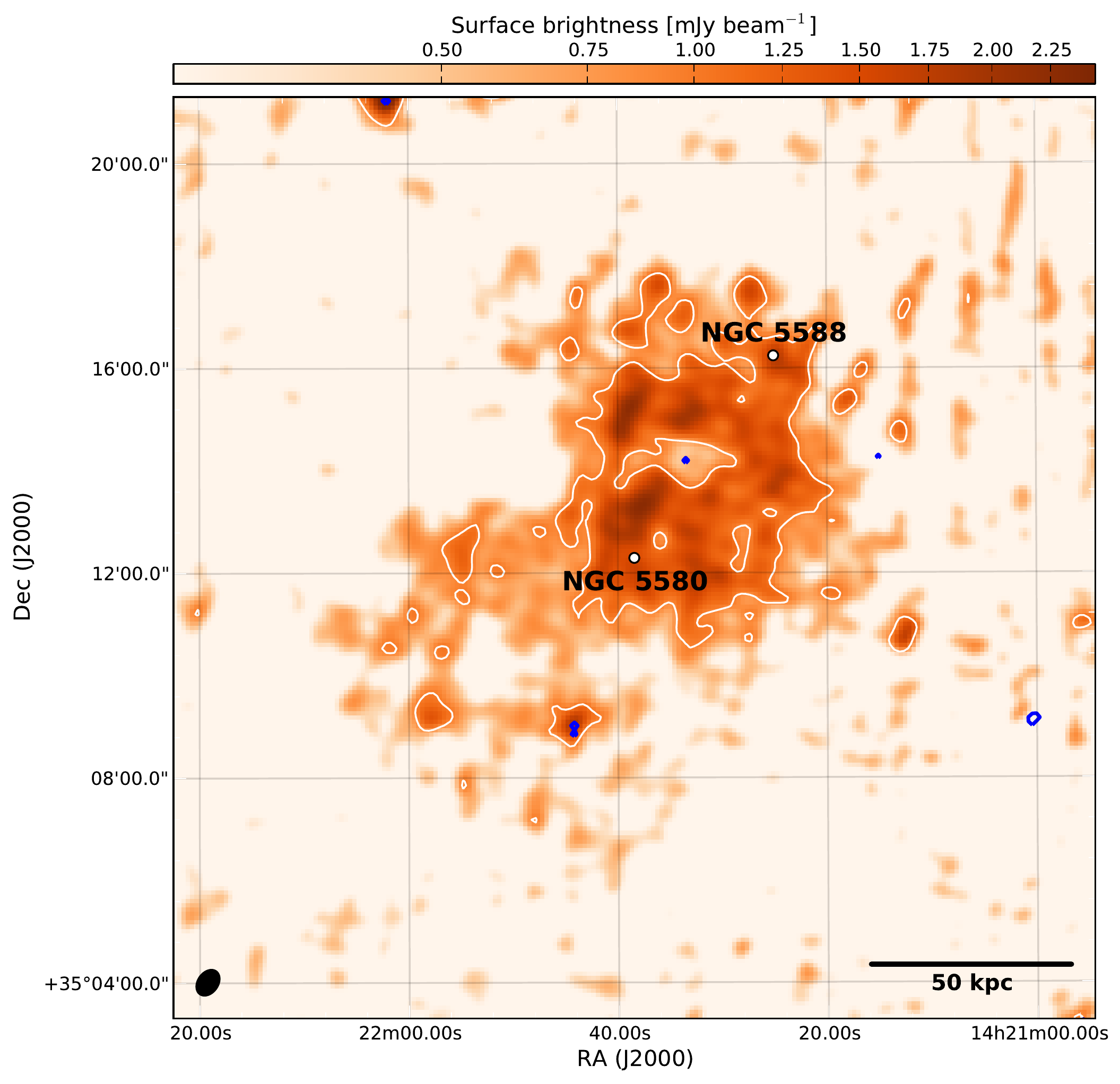}\label{fig:610_LR}}
\caption{Deep \gmrt{} images. White contours show the $3\sigma$ level (values in Table~\ref{tab:campaign}). Blue contours show the locations of subtracted point sources.}
\end{figure*}

Together with these new observations we obtained four more images of the target using archival data. \cite{Williams2013} used the GMRT to perform a 150 MHz survey, with one of their mosaic pointings almost centered on our target. The image obtained with their observation is shown in Fig.~\ref{fig:150_LR}, where the point sources were subtracted and data tapered using the same parameter as for the other \gmrt{} images to enhance the extended emission.

\begin{figure}
\centering
\subfloat[\wsrt{} (WENSS) 327 MHz.]{\includegraphics[width=.85\columnwidth]{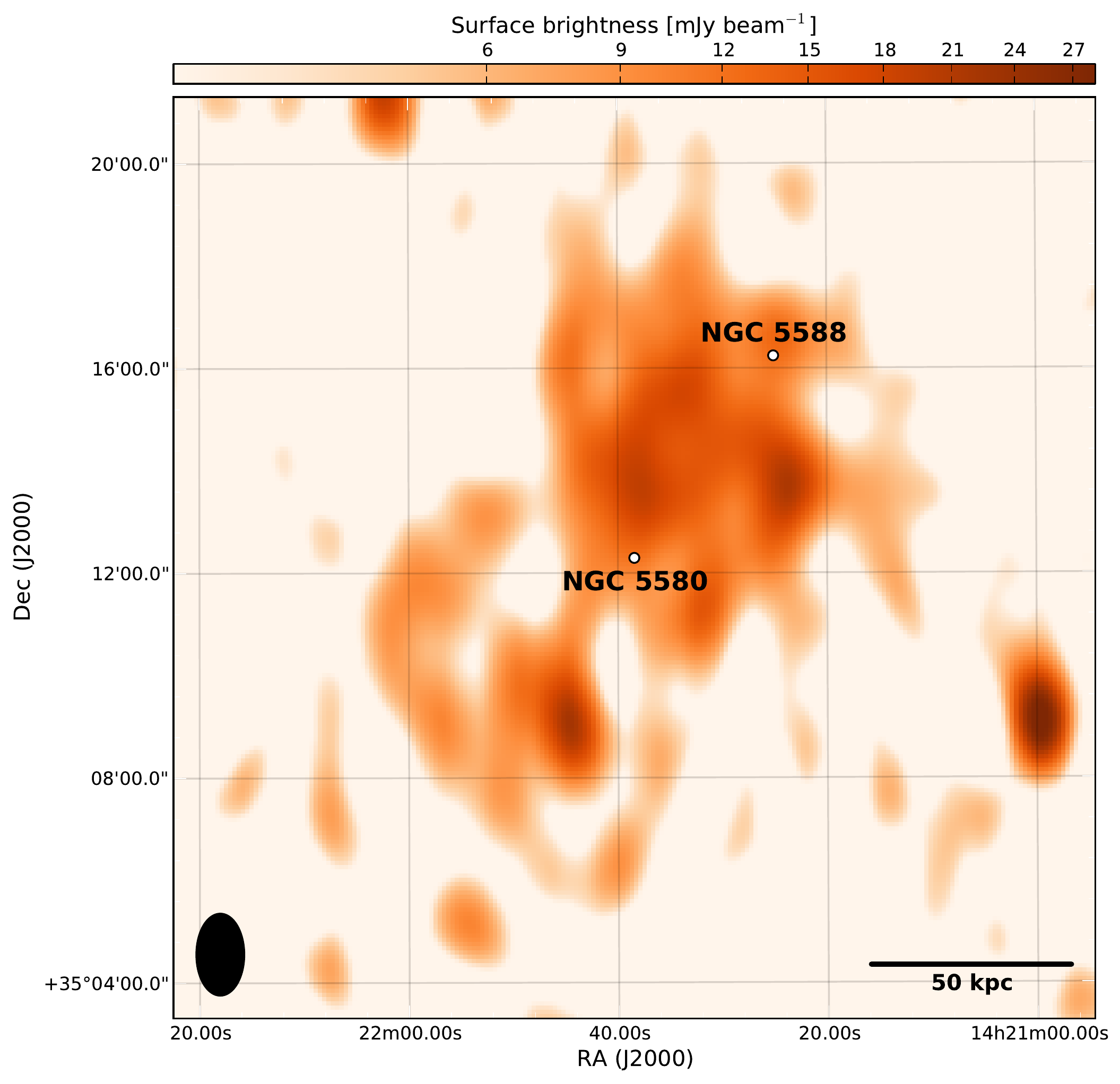}\label{fig:wenss}}\\
\subfloat[\vla{} (snapshot) 1400 MHz.]{\includegraphics[width=.85\columnwidth]{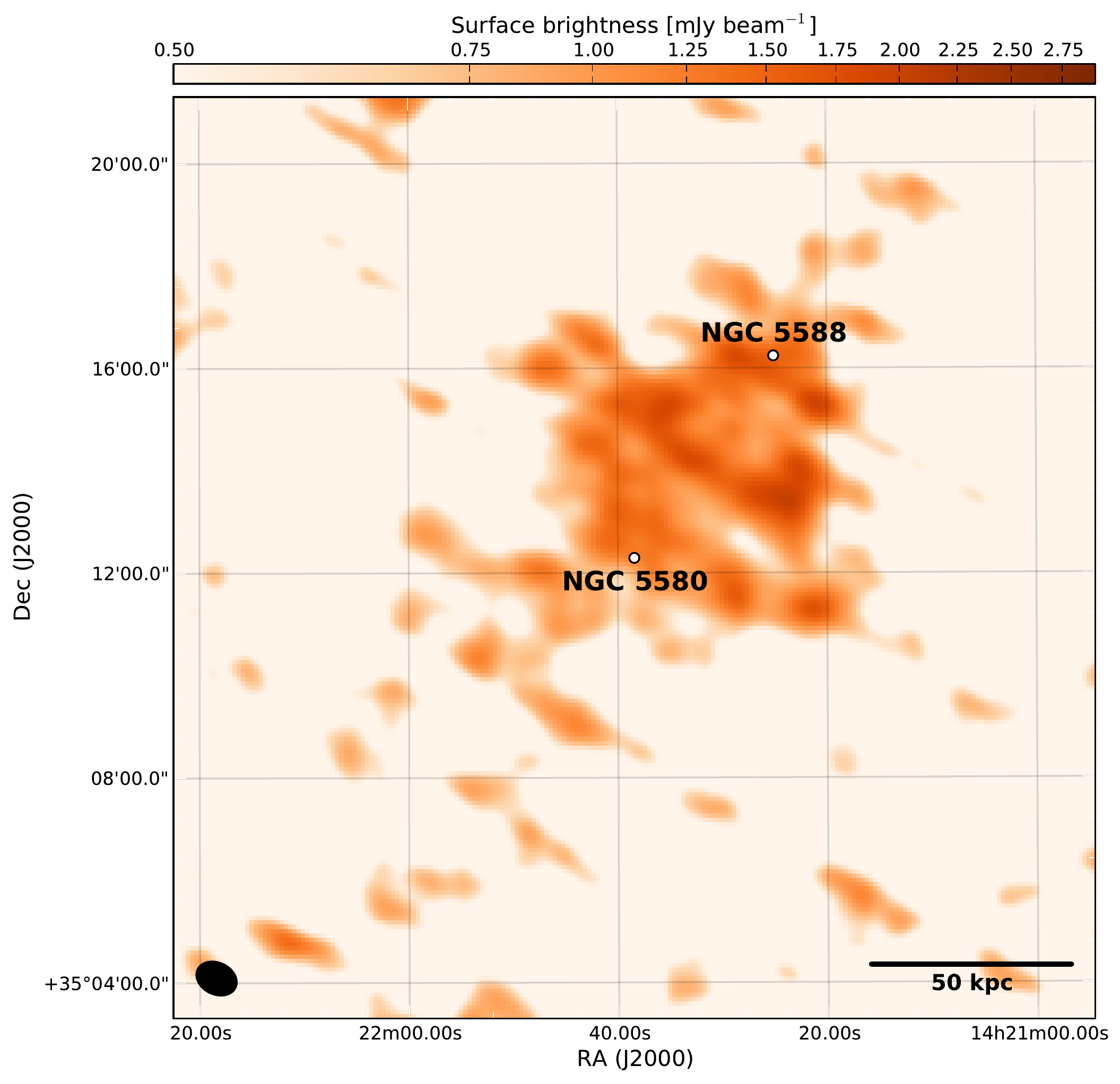}\label{fig:1400}}\\
\subfloat[\vla{} (NVSS) 1400 MHz.]{\includegraphics[width=.85\columnwidth]{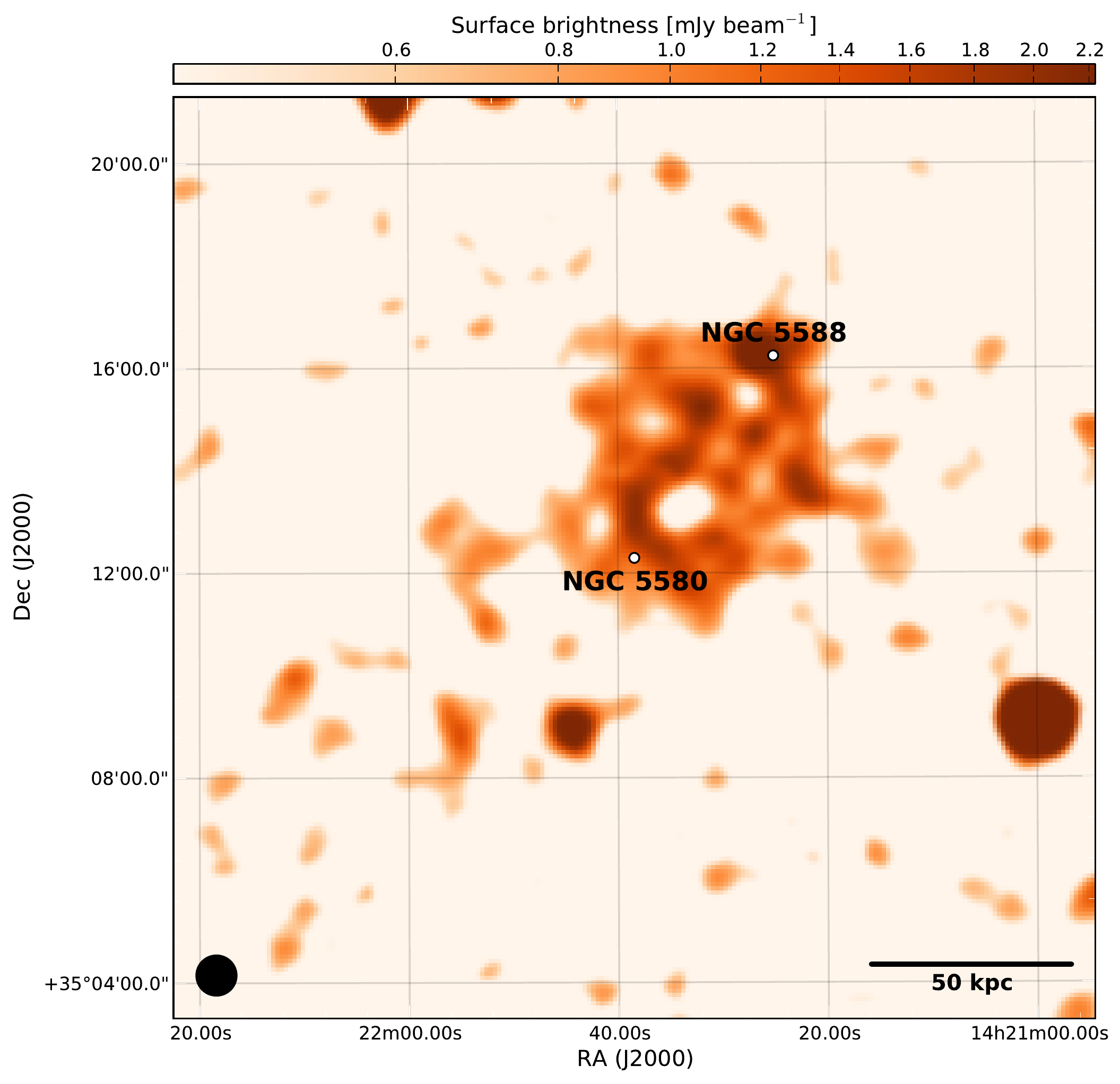}\label{fig:nvss}}
\caption{Snapshot/survey images obtained from archival data.}\label{fig:surveys}
\end{figure}

Searching the Very Large Array (\vla{}\footnote{\vla{}: \url{http://www.vla.nrao.edu}}) archive we found a 7-minute snapshot observation at 1400 MHz. Data were taken in 1997 with the ``old'' \vla{} in C-configuration. Although the target is well visible after tapering the data (see Fig. \ref{fig:1400}), due to the poor $uv$-coverage the fidelity of the image is limited. Finally, the target is visible both in the WENSS survey (Fig.~\ref{fig:wenss}) and the NVSS survey (Fig.~\ref{fig:nvss}).

\begin{table*}
\begin{center}
\caption{List of radio maps used in this paper.}
\label{tab:campaign}
\begin{tabular}{ccccccccc}
\hline
Telescope & Frequency & Resolution$^1$ & Largest scale$^2$ & Prim. beam & Obs. Time & RMS           & Figure & Reference \bigstrut[t]\\
          & [MHz]     &                & [armin]           & FWHM [deg] & [min]     & [~\mjybeam{}] &        & \bigstrut[b]\\
\hline
\gmrt{}         & 153  & \beam{25}{25} & 68 & 3.1 & 205$^3$     & 6.5 & \ref{fig:150_LR} & \cite{Williams2013} \bigstrut[t]\\
\gmrt{}         & 235  & \beam{41}{23} & 44 & 2.0 & $7\times60$ & 1.5 & \ref{fig:240_LR} & This work\\
\gmrt{}         & 323  & \beam{30}{26} & 32 & 1.4 & $7\times60$ & 0.4 & \ref{fig:330_LR} & This work\\
\wsrt{} (WENSS) & 327  & \beam{94}{54} & 21 & 2.7 & 36$^3$      & 3.1 & \ref{fig:wenss}  & \cite{Rengelink1997}\\
\gmrt{}         & 612  & \beam{31}{22} & 17 & 0.7 & $7\times60$ & 0.4 & \ref{fig:610_LR} & This work\\
\vla{}          & 1400 & \beam{47}{35} & 16 & 0.5 & 7           & 0.6 & \ref{fig:1400}   & This work\\
\vla{} (NVSS)   & 1400 & \beam{45}{45} & 16 & 0.5 & $<1^3$      & 0.5 & \ref{fig:nvss}   & \cite{Condon1998} \bigstrut[b]\\
\hline
\end{tabular} 
\end{center}
{$^1$ Beam major $\times$ minor axis. $^2$ The largest scale the instrument can detect, our target has an extension $\lesssim 8$\arcmin. $^3$ As part of a mosaic.
\par}
\end{table*}

All the images were rescaled to match the \cite{Perley2013} flux scale. For the four \gmrt{} datasets, the $uv$-ranges were cut in order to be equal before every imaging step. This procedure minimizes the systematic errors in radio-interferometric spectral index calculations of extended sources due to the under-sampling of the large-scale structures at the higher frequencies. The image resolutions and RMS values are listed in table~\ref{tab:campaign}.

\section{Source morphology}\label{sec:images}

The best image in terms of fidelity is the 323 MHz map obtained with the \gmrt{}. Not surprisingly, this was one of the three 8-hour long observations and the only one for which we recorded both polarizations, although a polarimetric study has not been attempted. Therefore, in the following description of the source morphology we will refer to Fig.~\ref{fig:330_LR}, although most of the described structures are present in all the images. We note that some errors in the morphological structures are expected on extended sources obtained with a snapshot observation done with an interferometer. Therefore, the images shown in Fig.\ref{fig:surveys} must be treated with caution.

\begin{figure*}
\centering
\includegraphics[width=\textwidth]{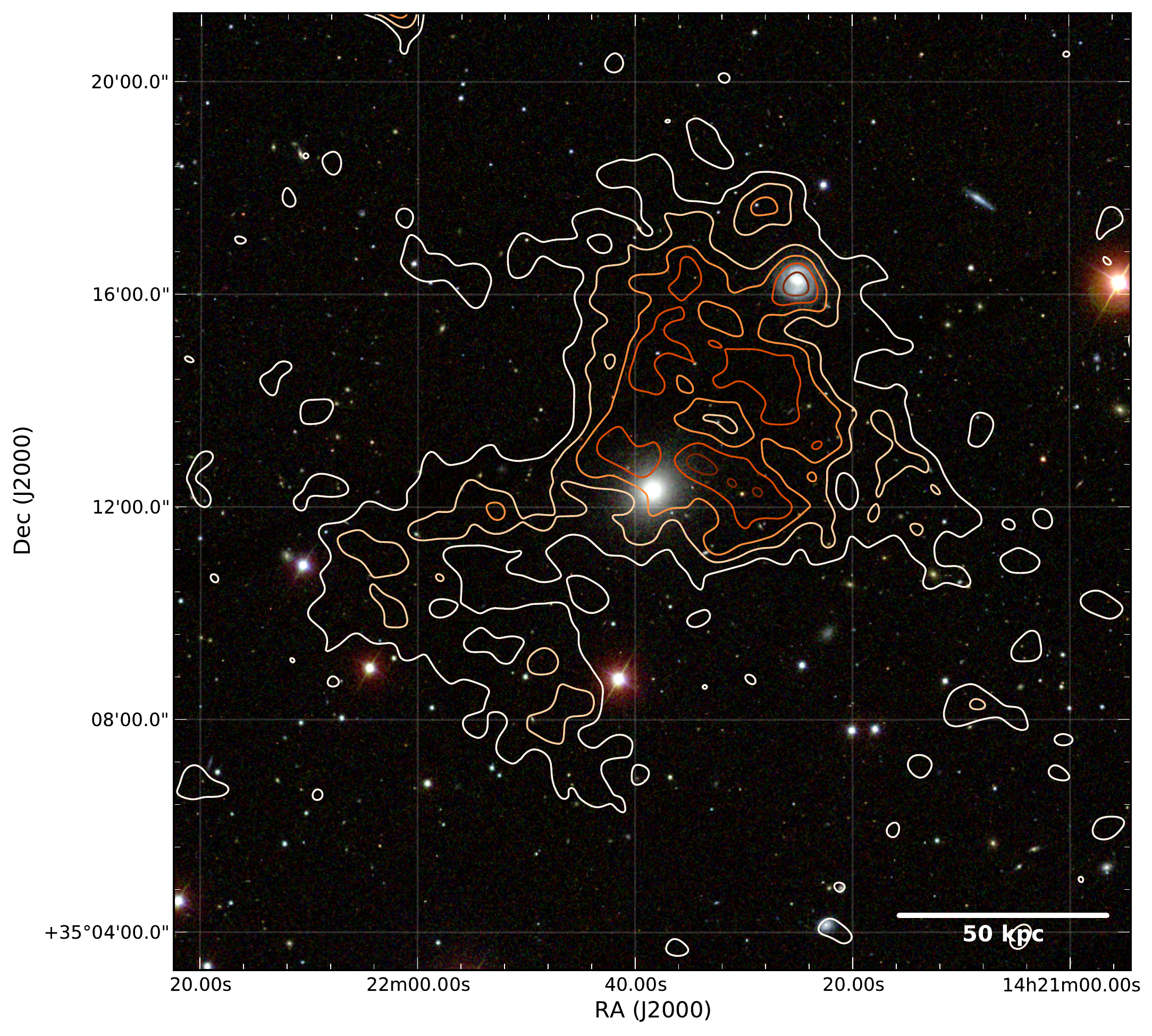}
\caption{SDSS image (g, r and i filters) of the target field with contours from Fig.~\ref{fig:330_LR}. Contours are at $\sigma\times(2,4,6,8,10)$, $\sigma = 350$~\mujybeam{} (values based on Fig.~\ref{fig:330_LR}, \gmrt{} at 323~MHz).}\label{fig:sdss}
\end{figure*}

The observed radio source appears to be a roundish \textit{bubble}, with a radius of around 50 kpc. The internal structures in the surface brightness distribution can double their values with respect to the mean in some zones of the sources and reduce it in others, close to the borders but also right at the centre.
Within the central bubble, the source appears to be brighter along three filaments: one in the East running North-South and another two running East-West which meeting the first filament at its base and midway up. An enhancement of emission is also present in the location of NGC~5588. These considerations underline the presence of a complex morphology which cannot be described by a smooth decrement of the flux moving from the centre to the outskirts nor by flow-like features. A fainter filament is connected to the South-East zone of the source (close to the position of NGC~5580) and extends towards East, bending after 50 kpc towards South-West. This filament is more visible in the low-frequency maps and it could be the border of a second \textit{bubble} with a lower surface brightness, which would be almost as large as the main one. The main, brightest \textit{bubble} has no clear boundaries. Its emission appears to fade-out at the borders generating some faint extensions like that visible in the South-East.

A comparison of the radio emission with an optical observation is presented in Fig.~\ref{fig:sdss}. The optical data is a composition of g, r and i filters from the SDSS\footnote{SDSS: \url{http://www.sdss.org/}} while contours are from the \gmrt{} observation at 323 MHz. The two large, bright galaxies are NGC~5580 (South-East) and NGC~5588 (North-West) and they are in projection located at the edges of the main \textit{bubble}. The first is a lenticular galaxy while the second is a barred-spiral. None of them have confirmed nuclear activity nor signs of recent interaction in the optical images. Superimposing the radio contour on the optical image, clear emission from NGC~5588 is detected. Since we have previously subtracted all the point-like sources, we believe this emission originates by star formation activity in the arms and the bar of the spiral galaxy rather than from a compact AGN. The point-like emission from NGC~5588 is in fact only 2.7~mJy (at 323 MHz), while the extended emission associated with the galaxy is about an order of magnitude stronger. NGC~5580 is instead located close to what could be the conjunction of the main \textit{bubble} with the second, fainter one.

Assuming a luminosity distance of 45.6 Mpc, the radio emission has a total power of $P_{323}\simeq9.4 \times 10^{22}$ W/Hz at 323 MHz.

\section{Spectral analysis}\label{sec:spidx}

\begin{figure}
\centering
\includegraphics[width=\columnwidth]{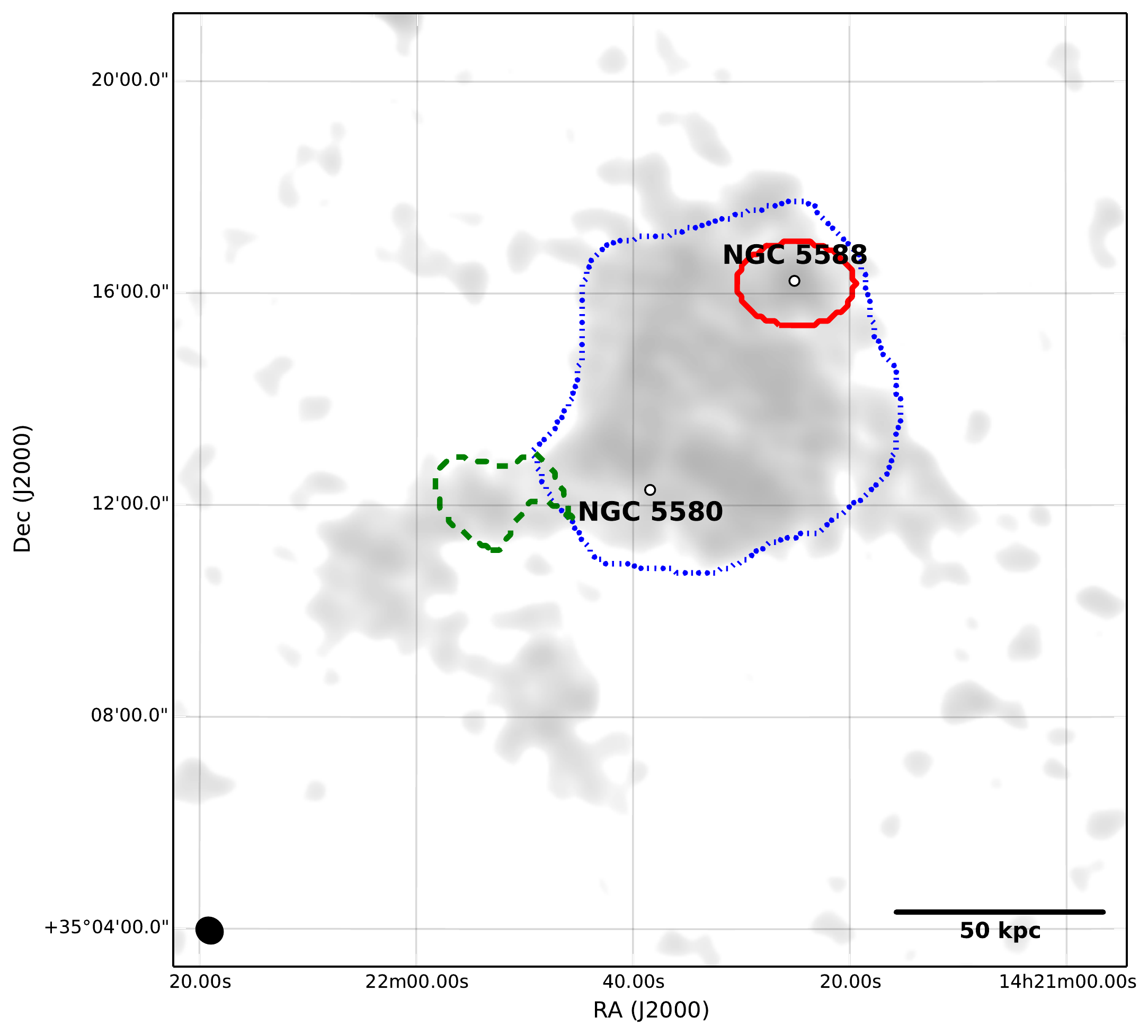}
\caption{Masks used to extract the spectral indexes in Fig.~\ref{fig:sed-zone1}, \ref{fig:sed-zone2} and \ref{fig:sed-NGC5588}. Blue (dotted-line): zone 1, the brightest lobe; green (dashed-line): zone 2, the fainter lobe; red (solid-line): NGC~5588. Each zone has been selected considering only values above $1 \sigma$ in all \gmrt{} maps.}\label{fig:masks}
\end{figure}

\begin{figure*}
\centering
\subfloat[Zone 1]{\includegraphics[width=.66\columnwidth]{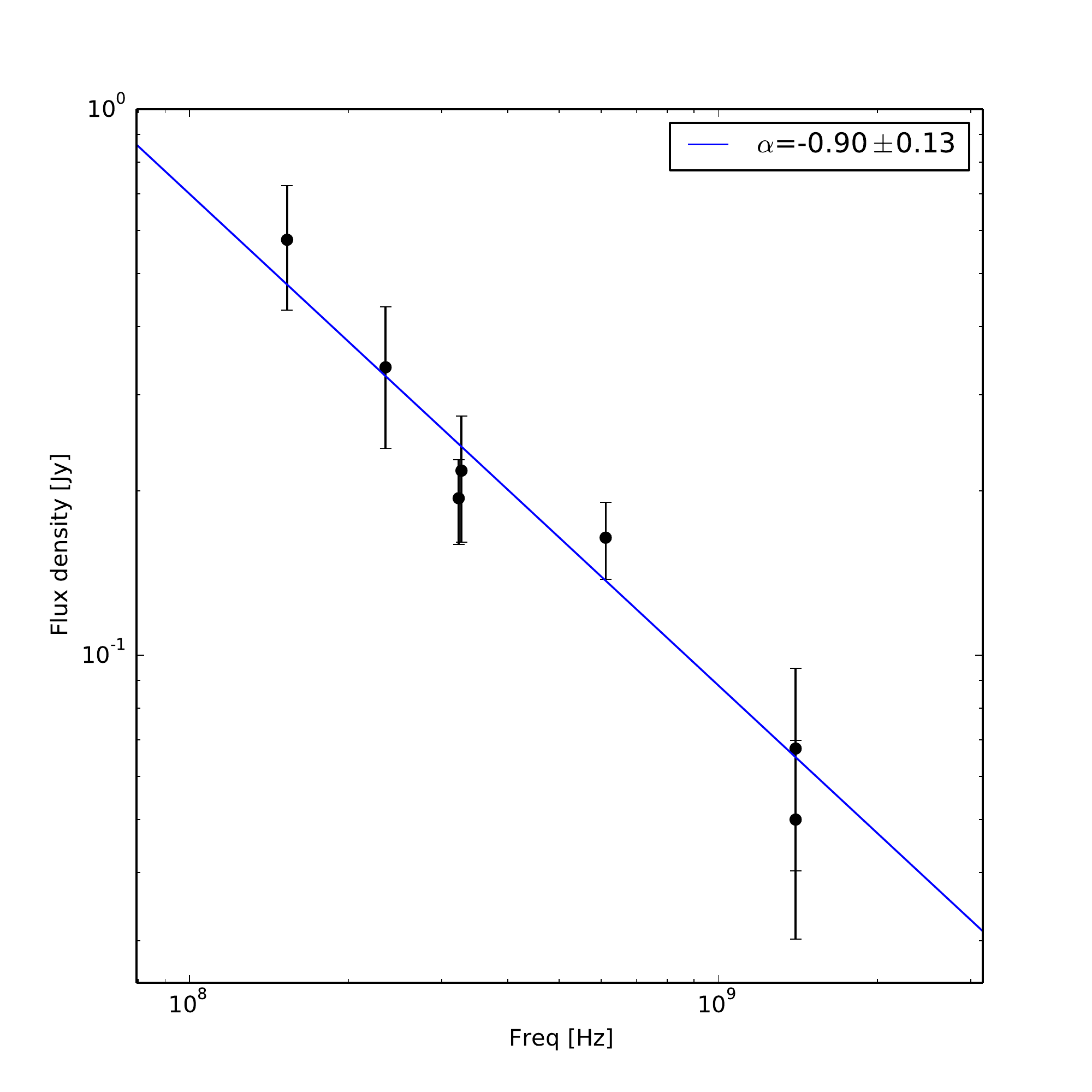}\label{fig:sed-zone1}}
\subfloat[Zone 2]{\includegraphics[width=.66\columnwidth]{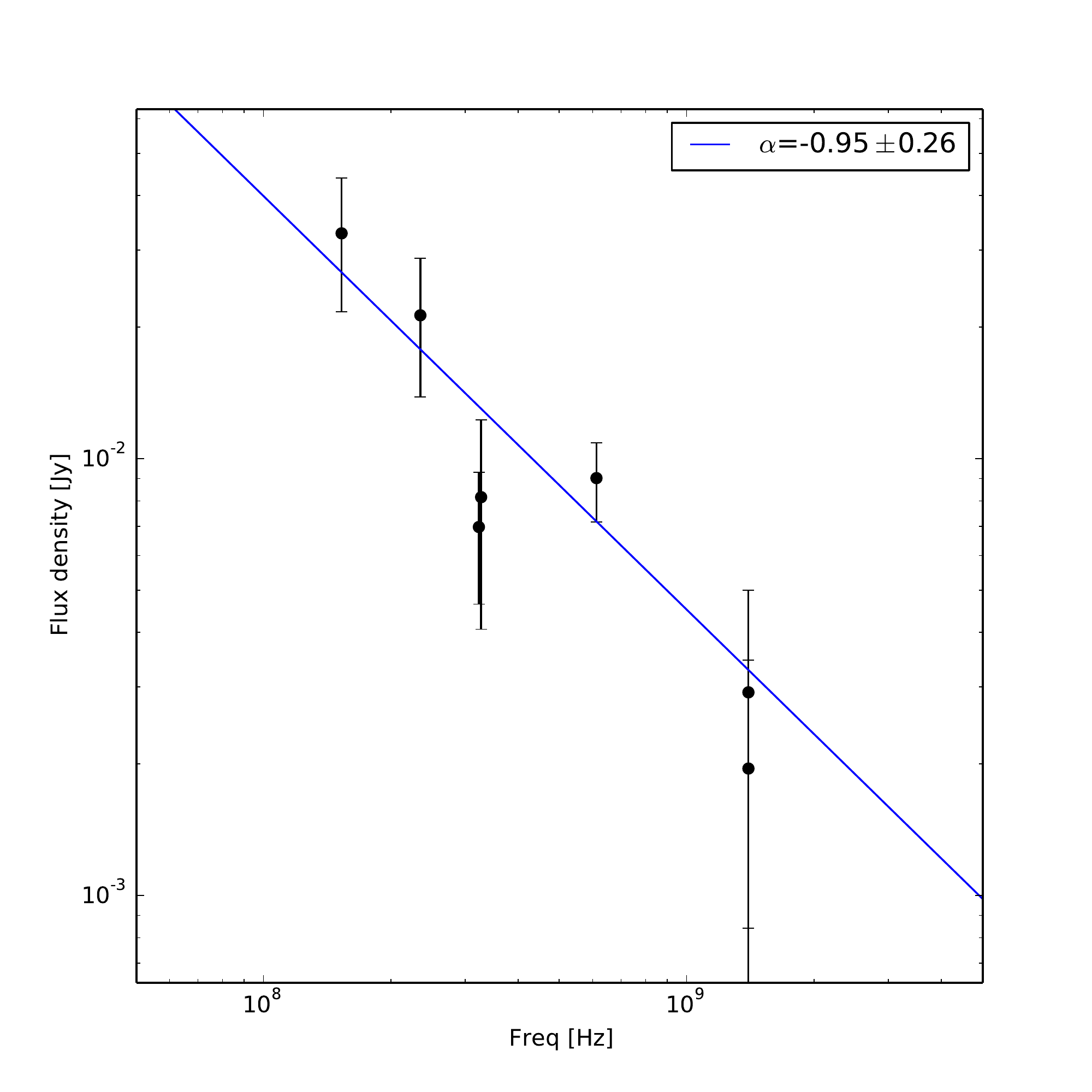}\label{fig:sed-zone2}}
\subfloat[NGC5588]{\includegraphics[width=.66\columnwidth]{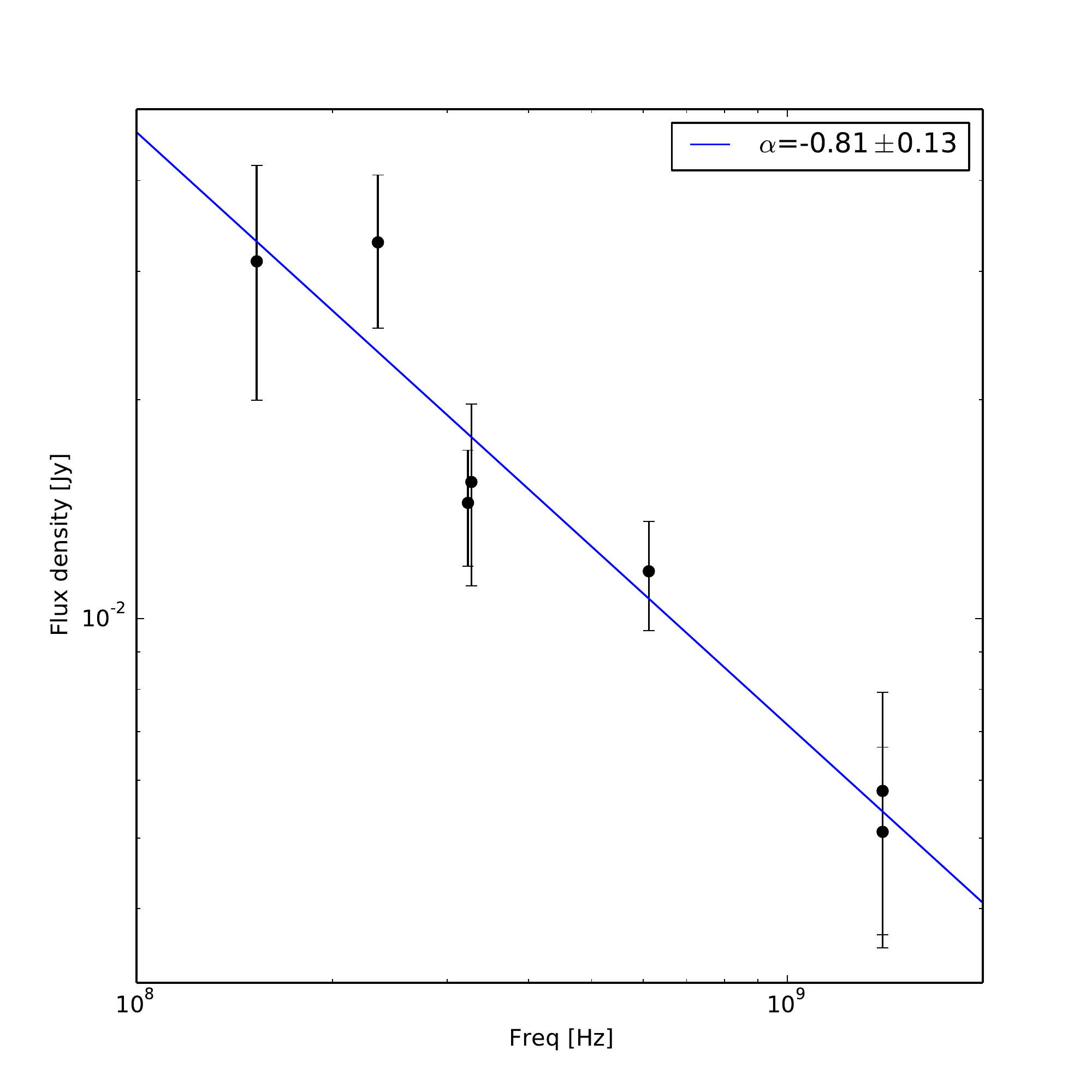}\label{fig:sed-NGC5588}}
\caption{Spectral energy distribution for different zones shown in Fig.~\ref{fig:masks}. Error bars are calculated as explained in the text (Sec.~\ref{sec:spidx}). Blue line is a linear regression.}
\label{sed}
\end{figure*}

We used the four deep \gmrt{} images, the image from the archival data at 1400 MHz and the two images from WENSS and NVSS to extract the global spectral index value\footnote{We use the spectral index definition: $S_{\nu} \propto \nu^{\alpha}$.} of the source. Firstly, we divided the source into zones, one around the brightest \textit{bubble} (zone~1), one around the South-East tail (zone~2) and a third one encompassing only NGC~5588. In the computation of every spectrum only pixels above $1\sigma$ in \textit{every} map considered were taken into account. Even if $1 \sigma$ is a low cut, we trust that the emission is real because it is detected at the same time in all the seven images. This cut slightly bias our global spectra analysis towards flatter spectra. This because our noisiest maps are at the higher frequencies, therefore we lose regions where data are present in the lower-frequency maps but are under the $1\sigma$-level at the higher-frequencies because of the spectral steepness. Given this constraint, the derived masks are shown in Fig.~\ref{fig:masks}. The error of the flux value $S_\nu$ of the extended zones has been calculated using
\begin{equation}
 \sigma_{S} = \sqrt{ \left(\sigma \sqrt{N_{\rm beam}}\,\right)^{2} + \left( S_\nu \times \delta_{\rm cal} \right)^{2} },
\end{equation}
where $\sigma$ is the RMS of the map, $N_{\rm beam}$ is the number of beam covering the extended region and $\delta_{\rm cal}$ is the flux calibration error that we conservatively assumed to be 10\%.
The spectra of these three zones are show in Fig.~\ref{fig:sed-zone1}, \ref{fig:sed-zone2}, and \ref{fig:sed-NGC5588}.

All extracted spectra can be fairly well approximated with a linear spectral energy distribution (SED) in this comparatively small frequency range. The slope of the averaged SED for zone~1 is $\alpha = -0.90\pm0.13$, while it becomes steeper, although compatible with the value of zone 1, moving to the tail (zone~2, $\alpha = -0.95\pm0.26$). Isolating the flux associated with the spiral galaxy NGC~5588 we extracted the spectrum shown in Fig.~\ref{fig:sed-NGC5588} ($\alpha = -0.81\pm0.13$), which is the combination of the extended emission due to the large \textit{bubble}, plus the emission due to star formation in NGC~5588 itself.

\begin{figure}
\centering
\captionsetup[subfigure]{labelformat=empty,aboveskip=0pt,parskip=0pt,farskip=-10pt,captionskip=0pt} 
\subfloat[]{\includegraphics[width=.9\columnwidth]{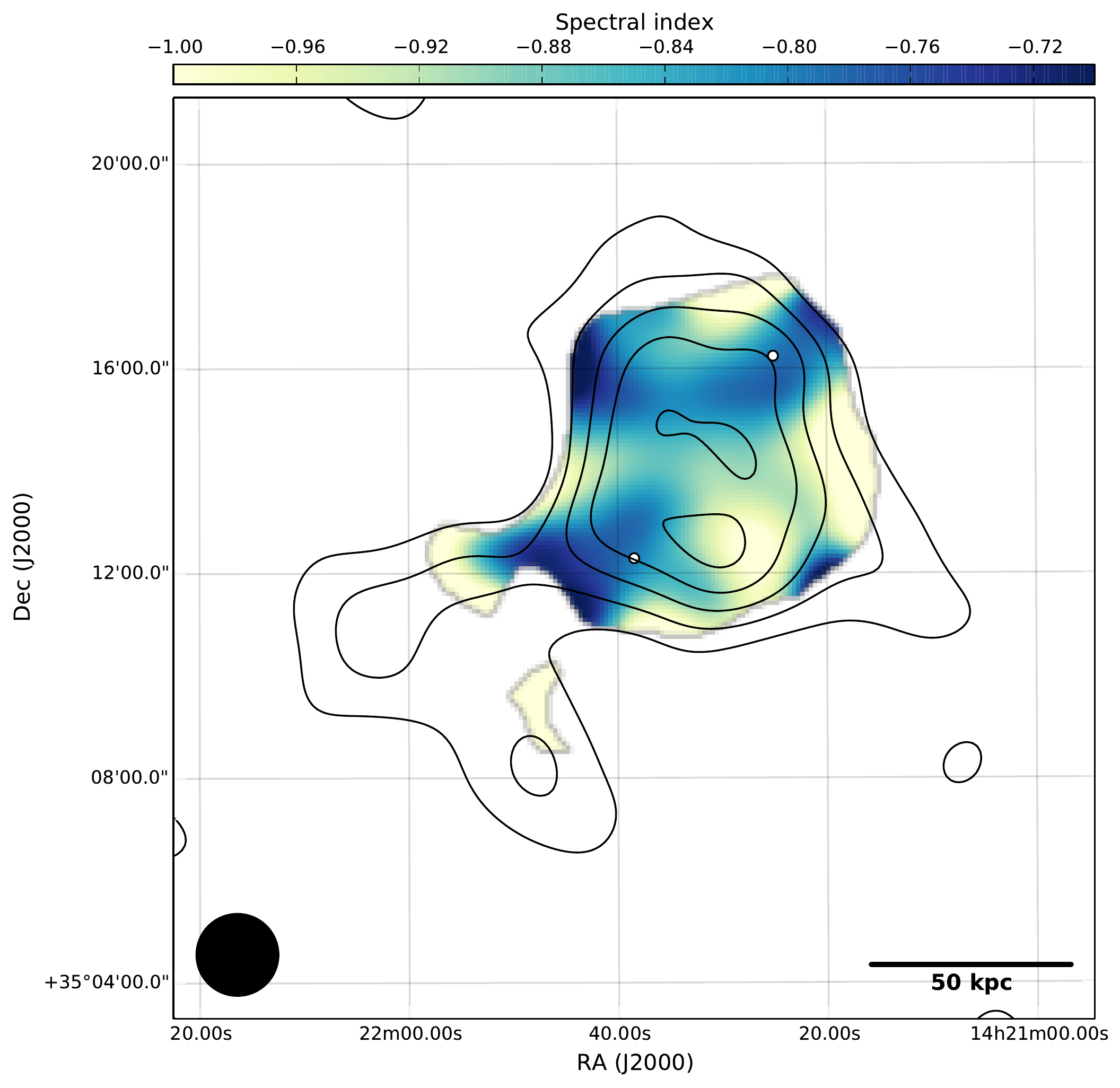}\label{fig:spidx-map}}\\
\subfloat[]{\includegraphics[width=.9\columnwidth]{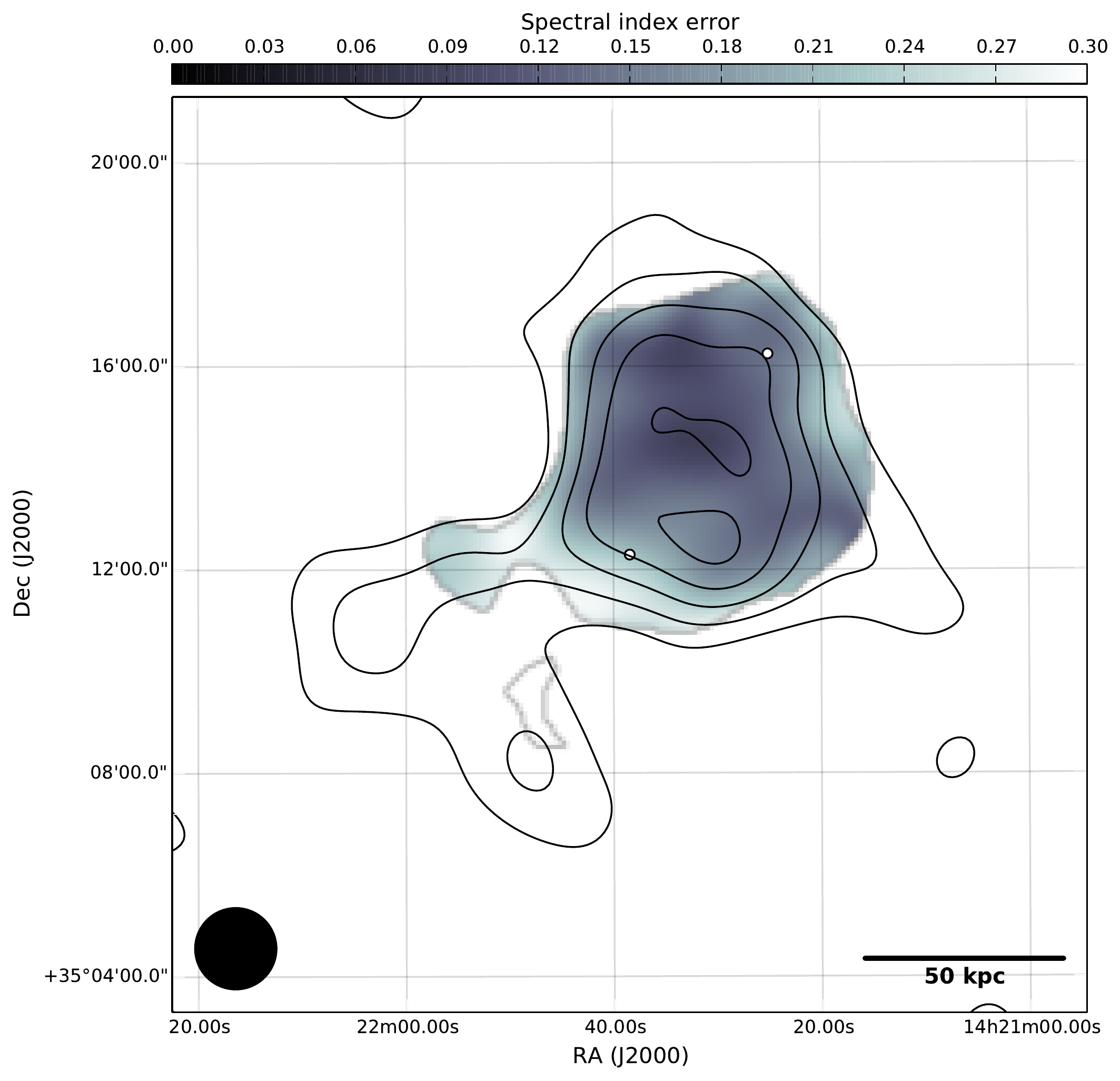}\label{fig:spidx-map-rms}}
\caption{\textit{Top:} spectral index map of the diffuse emission. \textit{Bottom:} spectral index error map. The maps resolution is 93\arcsec, contours are at $\sigma\times(1,2,3,4,5)$, $\sigma=2$~\mjybeam{} (values based on Fig.~\ref{fig:330_LR}, \gmrt{} at 323 MHz).}
\end{figure}

In Fig.~\ref{fig:spidx-map} we show a pixel-by-pixel spectral index map limited to all the pixels having a value larger then $1\sigma$ in \textit{every} map. The map does not show any clear linear or circular gradient. However, a sensible flattening in the spectral index is detected in the region around both galaxies. A flattening in the spectral index is usually interpreted as emission coming from younger population of CRe, as the higher-energetic CRe are depleted faster. When this happens and the particle energization has stopped, the spectra develop an exponential cut-off at the frequency $\nu_b$ which moves towards lower frequencies as the time passes following $\nu_b \propto t^{-2} B^{-3}$, where $t$ is the time from the CRe injection and $B$ is the magnetic field intensity.

We calculated the averaged equipartition magnetic field and the energy content of the brightest part of the source (zone~1 in Fig.~\ref{fig:masks}). We used the revised equipartition formula presented in \cite{Beck2005}. We assumed the emitting region to have a depth $l = 87.5$ kpc, which is the average diameter of the source. The ratio between CR electrons and protons has been set to $K_0 = 100$ and we used a spectral index of $\alpha = -0.87$. Given these parameters, the estimated magnetic field is $B_{\rm eq} = 2.5\ \mu\rm G$ and the total energy content is $E_{\rm tot} = 5.1 \times 10^{57}$ erg, which is typical for radio galaxies \cite{Birzan2008}. As a final note, we underline that given the spectral steepness, which implies strong losses, the equipartition values should be regarded only as tentative estimates.

\section{Discussion}\label{sec:discussion}

Although we obtained several images and the spectral information of our target, it is still hard to ultimately classify it.

\begin{figure}
\centering
\includegraphics[width=\columnwidth]{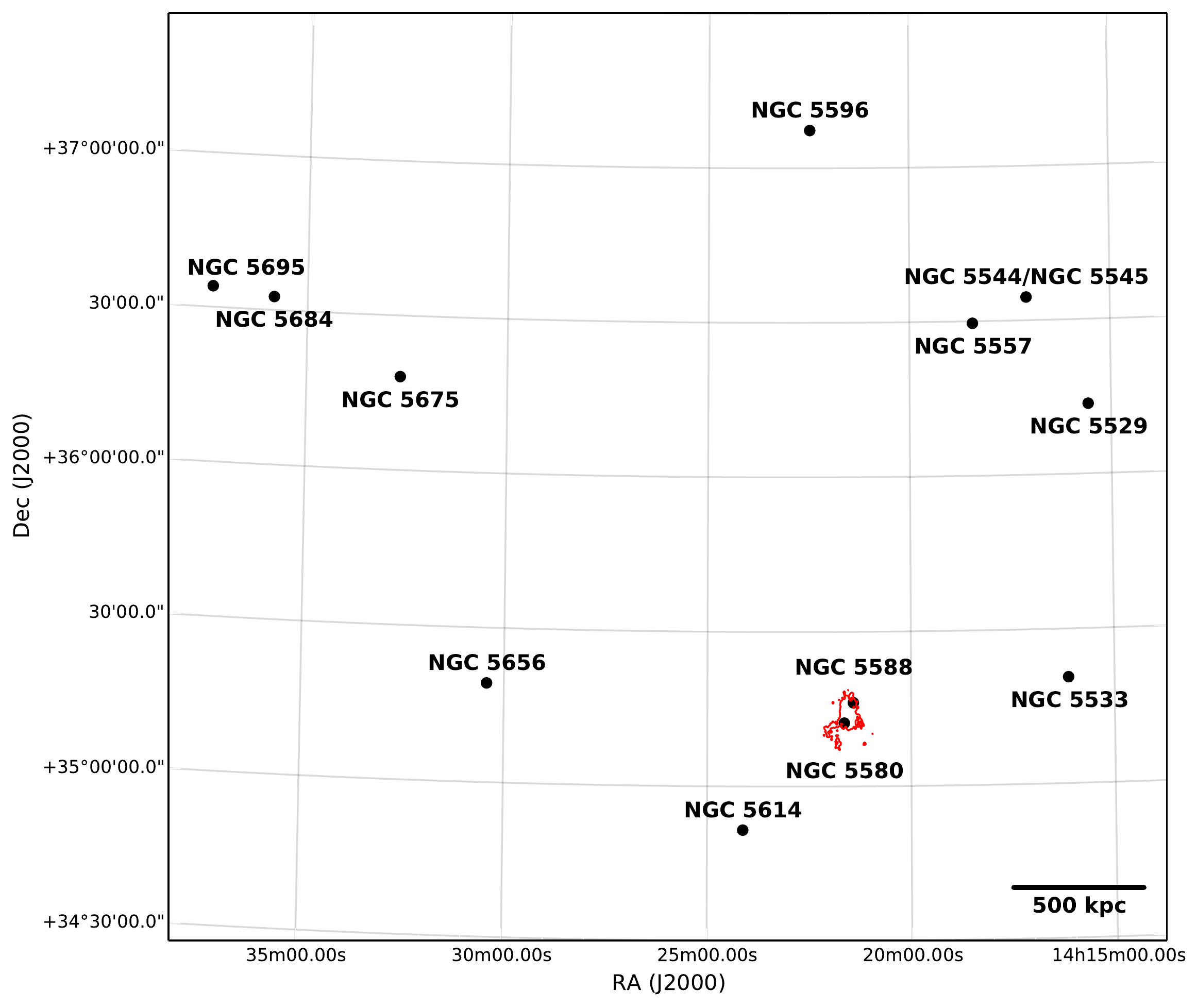}
\caption{Location of the galaxies which belong to the galaxy group G~141 \citep{Geller1983}. The emission surrounding NGC~5580 and NGC~5588 is shown as a red contour.}\label{fig:galaxygroup}
\end{figure}

A first possibility is that the emission is a down-scaled version of the Mpc-scale radio haloes present at the centre of merging galaxy clusters. This classification was first proposed by \cite{Delain2006}. As also noticed by the authors, no sign of high X-ray emission is visible in the \rosat{} survey. Although the exposure time is low ($<1$ ks), this is enough to put a constraint on the X-ray luminosity of $\log L_X < 41.2$ erg/s. This puts the source much below the known radio--X-ray correlation for radio haloes \citep[e.g.,][and references and therein]{Feretti2012}. There is evidence of radio haloes more radio-luminous than predicted \citep[e.g., Abell 523;][]{Giovannini2011, Farnsworth2013}. These powerful radio halos associated with low-density environments could be either young halos, or be related to clusters at a specific time of the merger event which triggered the halo formation (namely when particle acceleration processes are at a higher efficiency). However, these cases lie one order of magnitude above the radio--X-ray correlation, while the object of this paper is at least two further orders of magnitude brighter in radio then what is predicted. This consideration would be true even if the emission comes from an unknown background galaxy cluster (assuming a negligible k-correction). Another difference compared to cluster radio halos is that the source is not centrally located in the potential well of the galaxy group (see Fig.~\ref{fig:galaxygroup}). Finally, the source global spectral index is flatter compared with normal radio halo spectral indexes which at low frequencies are usually in the range $-1.2$ to $-1.8$ \citep{Feretti2012}.

A second possibility is that the emission is an extreme case of what are known as ``Taffy'' galaxies \citep{Hummel1986, Condon1993, Condon2002, Drzazga2011}. The angular projected distance between the two galaxies is 4\arcmin.78 = 62 kpc, which is larger than the typical distances of the ``Taffy'' galaxies ($\sim 10$ kpc). The proper distance of the two galaxies is unknown. When at the same distance, the two galaxies will have a differential line-of-sight velocity of $\sim189$ km/s based on the redshift difference, which is compatible with the velocity dispersion of the group which they belong \citep[461 km/s;][]{VanDriel2001}. However, the fainter/steeper filament towards the East would be hard to explain and the extended emission does not resemble a bridge connecting the two galaxies, like previous discovered examples of Taffy galaxies. Observation of emission lines as \hi{} can be used to estimate galaxies relative motion, distance and distortion due to interaction. This could help understanding whether the two galaxies had a recent close passage as requested by this scenario.

\begin{figure}
\centering
\includegraphics[width=0.95\columnwidth]{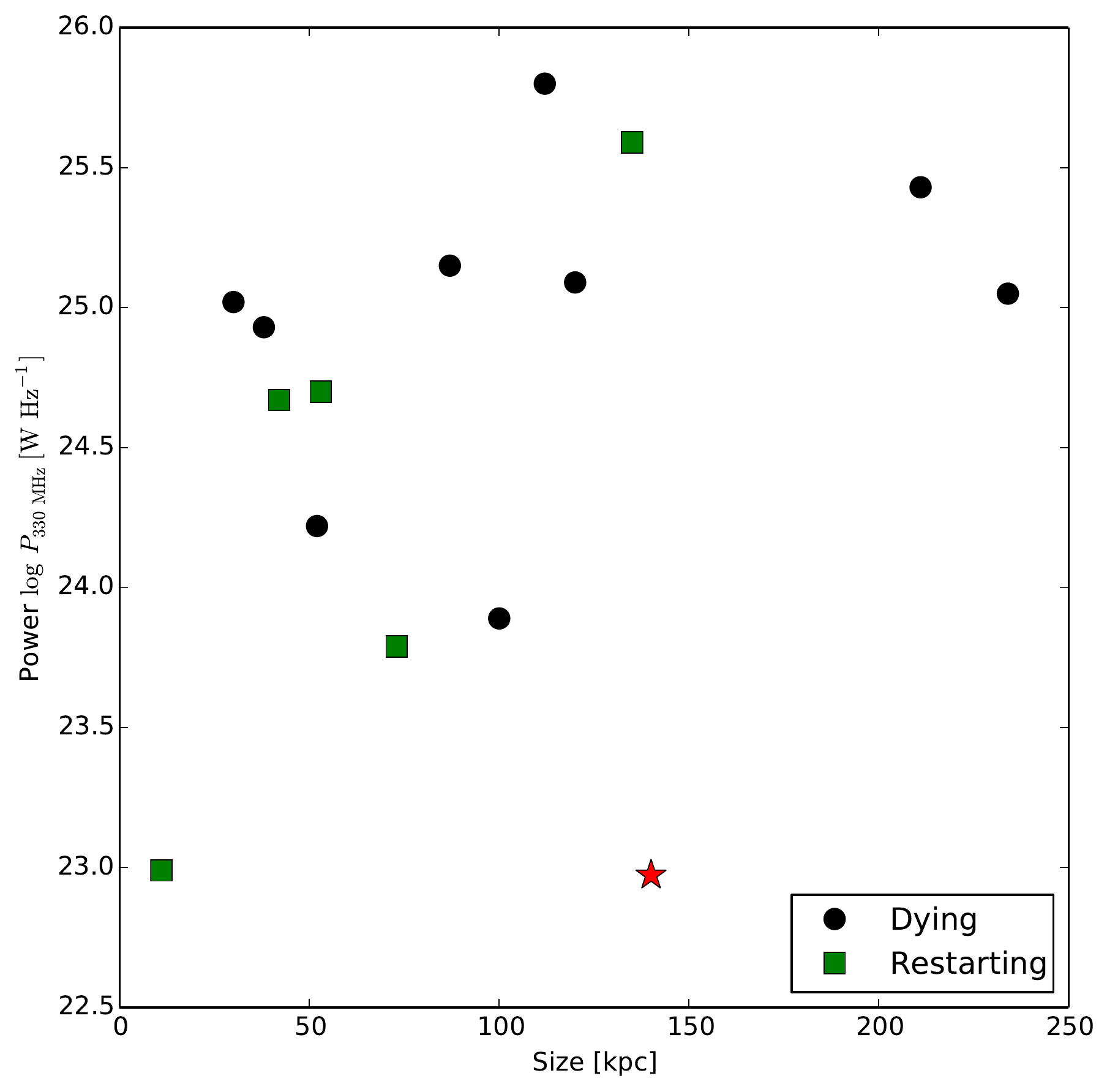}
\caption{ Linear size versus emitted radio power at 330 MHz for the sample of dying
 (black circles) and restarting (green squares) radio galaxies of
 \citet{Parma2007} and \citet{Murgia2011}. The location of NGC~5580 is marked
 as a red star. }\label{fig:size-pot}
\end{figure}

A third possibility is that we are observing a case of a dead radio galaxy. Since dead radio galaxies are usually found in dense galaxy clusters \citep{Murgia2011} and not in loose galaxy groups, this case is somewhat uncommon. However, major clues which brought us to favour this thesis are the source's brightness and spectral morphology. Without claiming complex projection effects, we could explain the two \textit{bubbles} as a relic of NGC~5580 AGN activity. However, in this case, the strong difference in surface brightness is uncommon (but not unique) and the presence of NGC~5588 just a chance alignment. Source spectra suggest a general steepening (i.e. ageing) as we move away from NGC~5580 which would be in line with the proposed scenario. The flattening in the spectral index map around NGC~5588 would be caused by the the dominating emission from the star formation in the spiral galaxy NGC~5588, over the dead-AGN radiation. The global spectral index of the source is not particularly steep, but this can be explained with the cut-off not having reached the GHz-frequency yet. Other examples of dying radio galaxies with a comparatively flat spectral index at low frequency are indeed known \citep[e.g., B2 1610+29;][]{Murgia2011} and are associated with a slow duty cycle where only the last burst of particle is still emitting at the lowest frequencies. Deep radio observations at lower and higher frequencies are necessary to further constrain the spectral shape of the source and confirm the picture. Compared to known samples of dying radio-galaxies \citep{Parma2007,Murgia2011}, NGC~5580 is faint for its extension (see Fig.~\ref{fig:size-pot}). This can be simply a selection effect since \cite{Parma2007} and \cite{Murgia2011} samples are selected from WENSS applying a flux cut at 30 mJy at 325~MHz and from the B2 survey applying a cut at 200 mJy at 408~MHz. At $z \simeq 0.1$, typical of the dying radio-galaxies in the two mentioned samples, NGC~5580 would have had a flux density $\lesssim 4$~mJy.

\section{Conclusions}\label{sec:concludions}

Using the \gmrt{} we observed the extended radio emission located in the galaxy group hosting NGC~5580 and NGC~5588. The radio emission is peculiar from, both, a morphological and spectral point of view. It shows one large \textit{bubble} with no apparent internal structure, encompassing the lenticular galaxy NGC~5580 and the spiral galaxy NGC~5588. The two galaxies sit at the edges of this extended emission, which has been previously classified as an uncommon example of radio halo in a galaxy group \citep{Delain2006}. The new observations revealed unambiguously the presence of a second \textit{bubble} of similar size as the first one, located to the South-East of NGC~5580. The spectral shape of the source shows a flattening of the SED around both galaxies.

The nature of this emission is probably related to a single past episode of AGN activity from NGC~5580, which would be able to explain both the source morphology and spectrum. In this case, the presence of NGC~5588 would be a coincidence and NGC~5580 would be a notable case of dead radio galaxy located outside a dense cluster. This implies that AGN do not require dense environments to undergo recurrent episodes of activity. If this picture will be confirmed by other examples, this would support a scheme where AGN are not preferentially triggered by mergers, in line with the recent findings of \cite{Karouzos2013}.

The fact that these type of sources are not often found outside of galaxy clusters may not be related to the intrinsic source population, since many radio sources are isolated and each of them must enter a dying phase. Indeed, \cite{Murgia2011} suggested that the probability to observe an isolated dying source is comparatively low because radio sources in under-dense environments may have a slower duty-cycle than those in dense environments and/or their lobes may fade faster given the poor confinement of the radio-emitting plasma. However, the observation of the fading phase of NGC~5580 shows that dying radio galaxies can be studied also in relatively isolated objects. Future large-scale low-frequency radio surveys, as those being carried on by the Low Frequency Array (LOFAR) and in the future by the Square Kilometre Array (SKA), will provide enough statistics to shed light on this intriguing topic.

\section*{Acknowledgements}
We would like to thank Ishwara Chandra C. H. and the staff of the \gmrt{} for the great help in setting up the observations. The \gmrt{} is run by the National Centre for Radio Astrophysics of the Tata Institute of Fundamental Research.\\
The National Radio Astronomy Observatory is a facility of the National Science Foundation operated under cooperative agreement by Associated Universities, Inc.\\
Funding for the SDSS and SDSS-II has been provided by the Alfred P. Sloan Foundation, the Participating Institutions, the National Science Foundation, the U.S. Department of Energy, the National Aeronautics and Space Administration, the Japanese Monbukagakusho, the Max Planck Society, and the Higher Education Funding Council for England.\\
A.B. and M.B acknowledge support by the research group FOR 1254 funded by the Deutsche Forschungsgemeinschaft: "Magnetisation of interstellar and intergalactic media:the prospects of low-frequency radio observations".

\bibliographystyle{mn2e}
\bibliography{NGC5580}
\bsp

\label{lastpage}

\end{document}